\documentclass[twoside]{ilcws08}
\usepackage[latin1]{inputenc}
\usepackage[dvips]{graphicx,epsfig,color}
\usepackage{wrapfig,rotating}
\usepackage{amssymb,amsmath,array}

\begin{document}

\title{Tile HCAL Test Beam Analysis:\\ Positron and Hadron Studies }

\author{Riccardo Fabbri
\vspace{.3cm}\\
on behalf of the CALICE Collaboration\\
FLC, DESY, Notkestrasse 85, 
22607 Hamburg, Germany\\
E-mail: Riccardo.Fabbri@desy.de\\
}

\maketitle

\begin{abstract}
The CALICE collaboration has constructed a hadronic sandwich calorimeter
prototype with 7608 scintillating plates, individually read out 
by multi-pixel silicon photomultipliers (SiPMs). For the first time ever 
the read out is performed using SiPMs on a large scale. Results of 
test beam operations with muon, positron and hadron beams at CERN are 
presented here, validating the feasibility of the novel SiPM technology. 
Results of the application of the particle flow approach in shower energy 
reconstruction are presented for the first time ever using real data.
\end{abstract}
%
\section{Introduction}
\vspace{-0.2cm}
\label{sec:intro}
The CALICE collaboration is performing calorimeter development aiming 
to fulfill the hardware and physics demands of the International 
Linear Collider physics program~\cite{ILC}. The ambitious required 
jet energy precision ($\approx 0.3/\sqrt{E(GeV)}$)
could be achieved with extremely segmented calorimeters using the particle 
flow approach (PFLOW)\cite{ILC}.

The CALICE tile hadron calorimeter prototype (HCAL) is a $38$ layer 
sampling calorimeter with $1$ m$^2$ lateral dimension, and total 
thickness of $5$ nuclear interaction lengths.
Each layer consists of $2$ cm thick steel absorber and a plane 
of $0.5$ cm thick plastic scintillator tiles. 
The tile sizes vary from $3$x$3$ cm$^2$ in the center of the layer, 
to $6$x$6$ cm$^2$ and $12$x$12$ cm$^2$ in the outer regions.
Each tile is coupled to a SiPM via a wavelength shifting fiber.

Together with the CALICE silicon-tungsten electromagnetic calorimeter 
(ECAL) and the CALICE tail-catcher and muon tracker (TCMT), the HCAL 
was exposed to muon, positron and hadron beams at the H$6$ test beam 
line at CERN in 2006 (partially instrumented with 23 layers) and 2007 
(fully instrumented with 38 layers, 7608 scintillating plates).
The results of the CERN data analysis are presented here. 
%
\section{Calorimeter calibration}
\label{sec:calib}
\vspace{-0.2cm}
The SiPM is a multi-pixel avalanche photodiode which provides a signal 
gain factor of $\approx 10^6$. The gain of each SiPM is monitored via 
dedicated 
measurements during test beam data taking, illuminating the device with 
low intensity LED light~\cite{CALICE}. 
During the extensive test beam operations at CERN a gain calibration 
efficiency of $\approx 97\%$ was observed, confirming the stable and 
high performance of SiPMs in a large scale calorimeter.  

The energy deposited by a particle in the scintillating plate is 
read by the SiPM, and converted into ADC units. The energy calibration
was performed measuring the response to the passage of minimum 
ionising particles (mip), using muons as mips~\cite{CALICE}. 
The muon signal deposited in a tile was fitted with a Gaussian convoluted 
with a Landau 
distribution, and the most probable value of the distribution was assumed 
to be one MIP unit, corresponding to $0.861$ MeV, as obtained by 
Monte Carlo simulations~\cite{CALICE}.
The systematical uncertainty on the energy calibration due to the 
intrinsic short-term operational variations of the detector properties 
was found to be $\approx 3\%$. In the analysis, the rejection of hits with 
energy below $0.5$ MIP results in a mip hit detection efficiency of 
about $93\%$~\cite{CALICE}.

\begin{figure}[t!]
   \begin{minipage}{8cm}
          \includegraphics[width=0.95\columnwidth, height=5.5cm]
                          {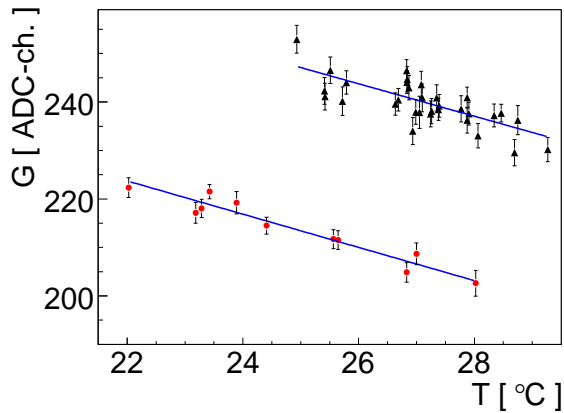}
   \end{minipage}
   \begin{minipage}{6.cm}
   \vspace{-0.9cm}
   \caption{Typical measured gain at different temperature values
            in an arbitrary calorimeter cell. The two data sets have
            been collected at $41.4$ V and $40.6$ V reverse voltage 
            values applied to the SiPM (filled triangles and circles, 
            respectively). A linear fit is performed to extract the linear
            temperature dependence of the gain, and the extracted
            functional form is used to correct the measured deposited energy
            for temperature effects.}
   \label{fig:GainAndTemp}
   \end{minipage}
   \vspace{-0.65cm}
\end{figure}

Due to the limited number of pixels and to the finite pixel recovery time
\mbox{($>100$ ns)}, SiPMs are non linear devices.
The reconstructed energy is corrected for non-linearity effects 
using response curves, measured individually for every SiPM~\cite{CALICE}.
SiPMs are temperature dependent devices. Therefore, a temperature 
monitoring system was implemented to measure and correct for the single 
pixel gain, Fig.~\ref{fig:GainAndTemp}, and for the mip amplitude 
temperature dependence. The slope of the temperature dependence of the 
gain $G$ and of the mip amplitude $A$, averaged over all channels,
was found to be $\frac{1}{G}\frac{d}{dT}G = -1.7\frac{\%}{K}$ 
and $\frac{1}{A}\frac{d}{dT}A = -3.7\frac{\%}{K}$~\cite{CALICE}, 
respectively.
%
%
\section{Analysis of positron data}
\label{sec:positrons}
\vspace{-0.2cm}
Being the description of underlying physics reasonably understood, 
Monte Carlo (MC) simulation of electromagnetic showers in the detector 
can be compared to the data to validate the calibration procedure, and 
to validate the detector effects introduced in the Monte Carlo simulation 
(digitisation).
The data shown here have been collected during 2007.

\begin{figure}[t!]
   \vspace{-0.15cm}
   \includegraphics[width=0.55\columnwidth, height=6cm]
        {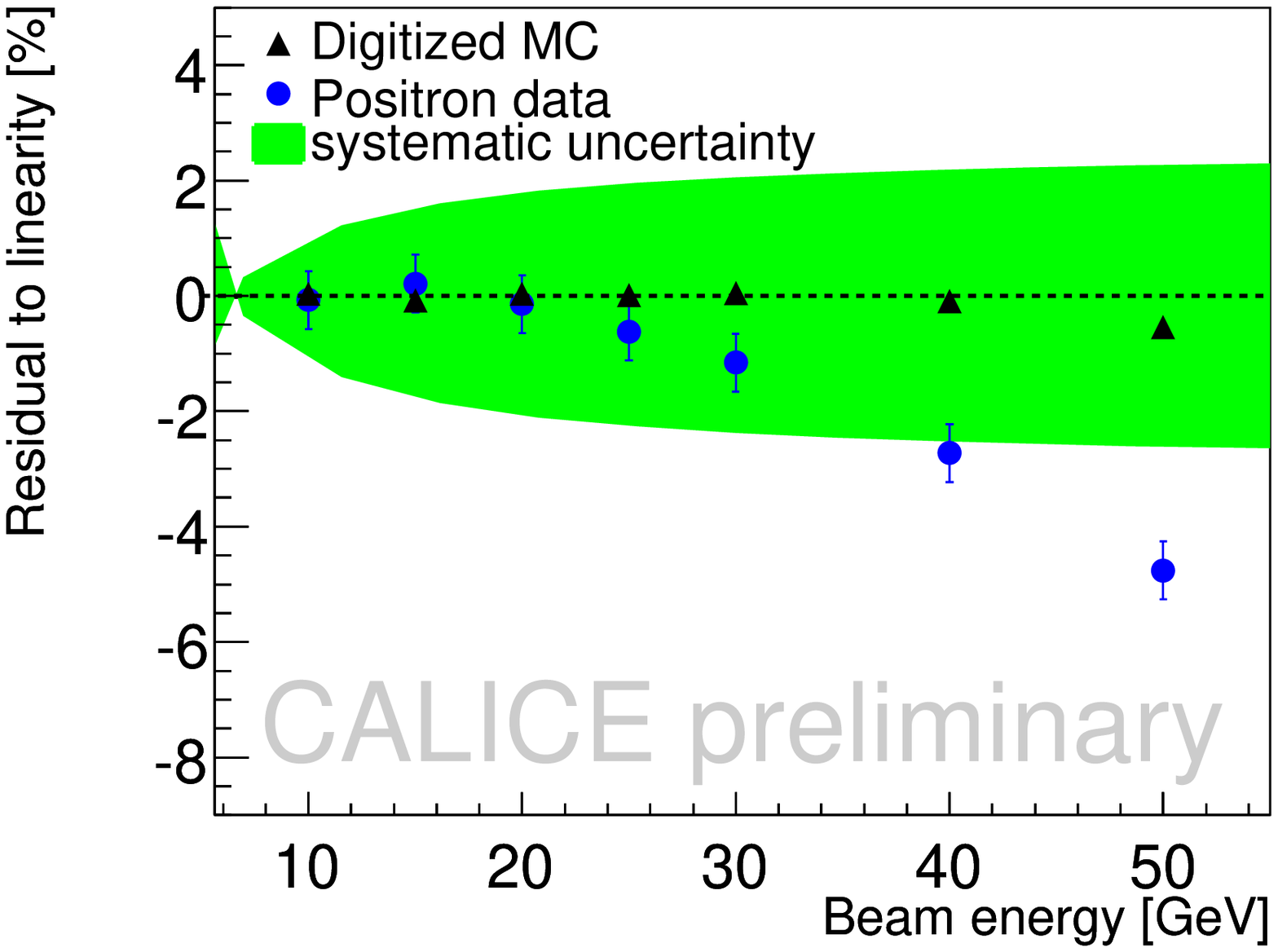}
   \hspace{-0.9cm}
   \includegraphics[width=0.55\columnwidth, height=6.cm]
        {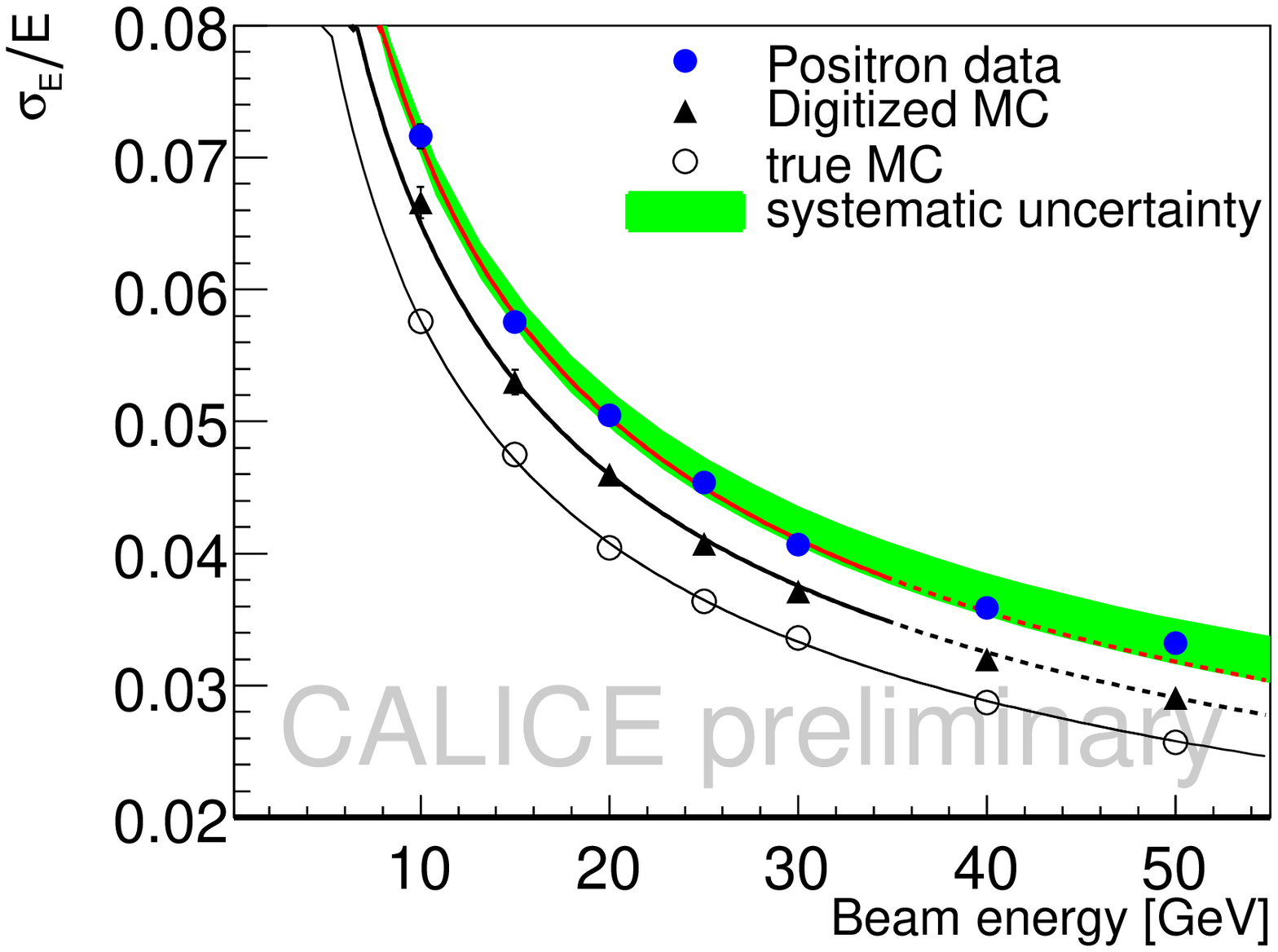}
   \vspace{-0.95cm}
   \caption{Residuals to the linear fit performed to 
            the detector response to electromagnetic showers 
            (left panel), and energy resolution to electromagnetic 
            showers (right panel) are shown for both the data and MC. 
            The largest contribution to the systematical uncertainty 
            is given by the calibration uncertainties on both the MIP 
            energy determination and the non-linearity effects correction.}
   \label{fig:EmLinearity}
   \vspace{-0.3cm}
\end{figure}

After applying all the calibrations described above, the detector 
response to electromagnetic showers is measured at different beam energy 
values. The residuals
to the linear fit to the data is shown in the left panel of 
Fig.~\ref{fig:EmLinearity}. Within both statistical and systematical 
uncertainties, the reconstructed response is linear up to $30$ GeV 
beam energy. Superimposed to the data, is also shown the MC prediction. 
The energy resolution is shown in the right panel of 
Fig.~\ref{fig:EmLinearity}, and compared with simulations with and without 
the inclusion of detector effects. The smearing effects included in the 
simulation improve the agreement, although the data have still systematically
larger values than what predicted by the simulations. 
This can be possibly understood considering that not all the calibration
uncertainties have been included in the MC yet. Their inclusion
should result in larger values of the simulated resolution. 
%
%
%
%
%
\begin{figure}[b!]
   \vspace{-0.7cm}
   \hspace{-0.3cm}
   \includegraphics[width=0.50\columnwidth, height=6cm]
        {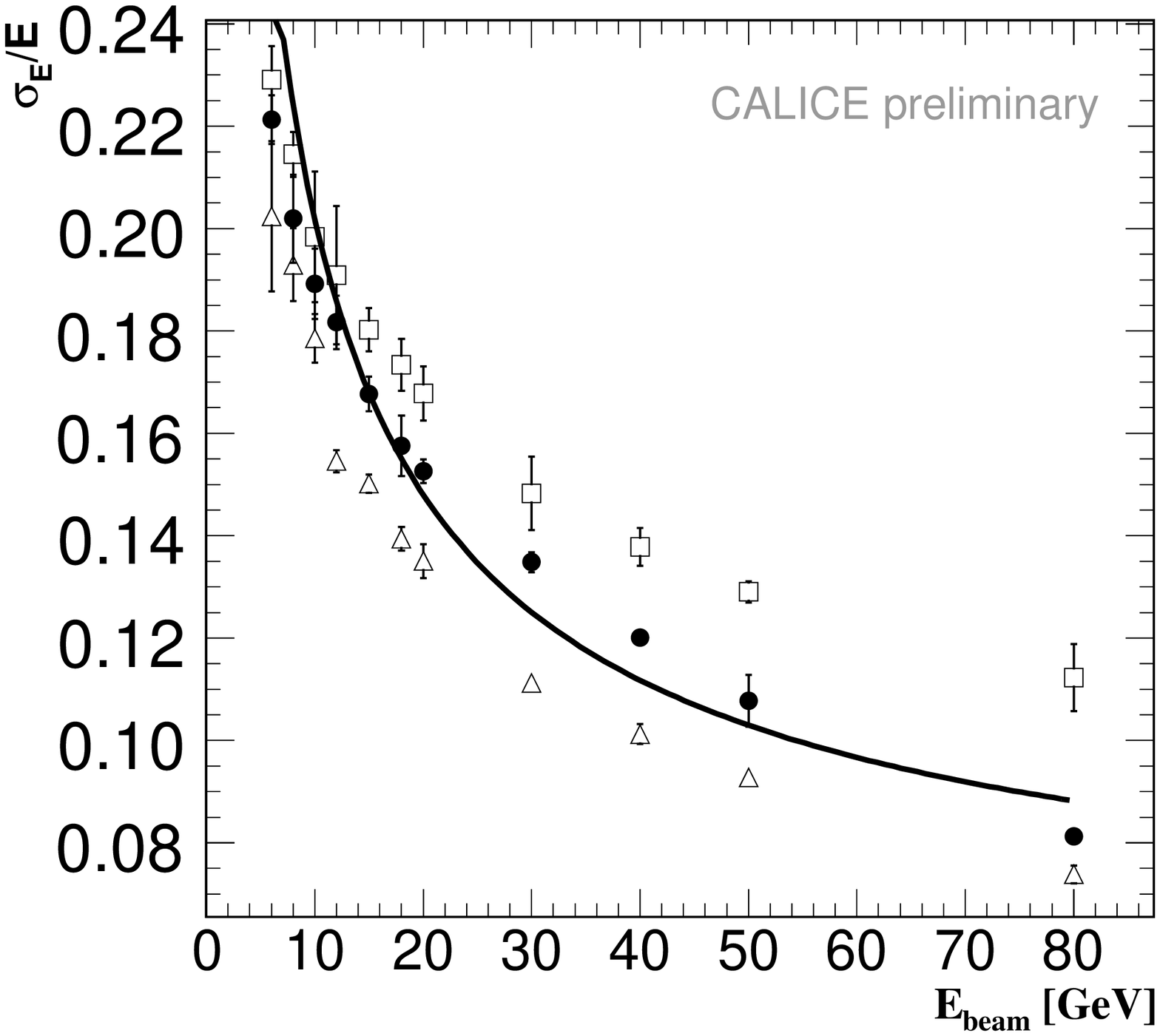}
   \hspace{-0.4cm} 
   \includegraphics[width=0.50\columnwidth, height=5.6cm ] 
        {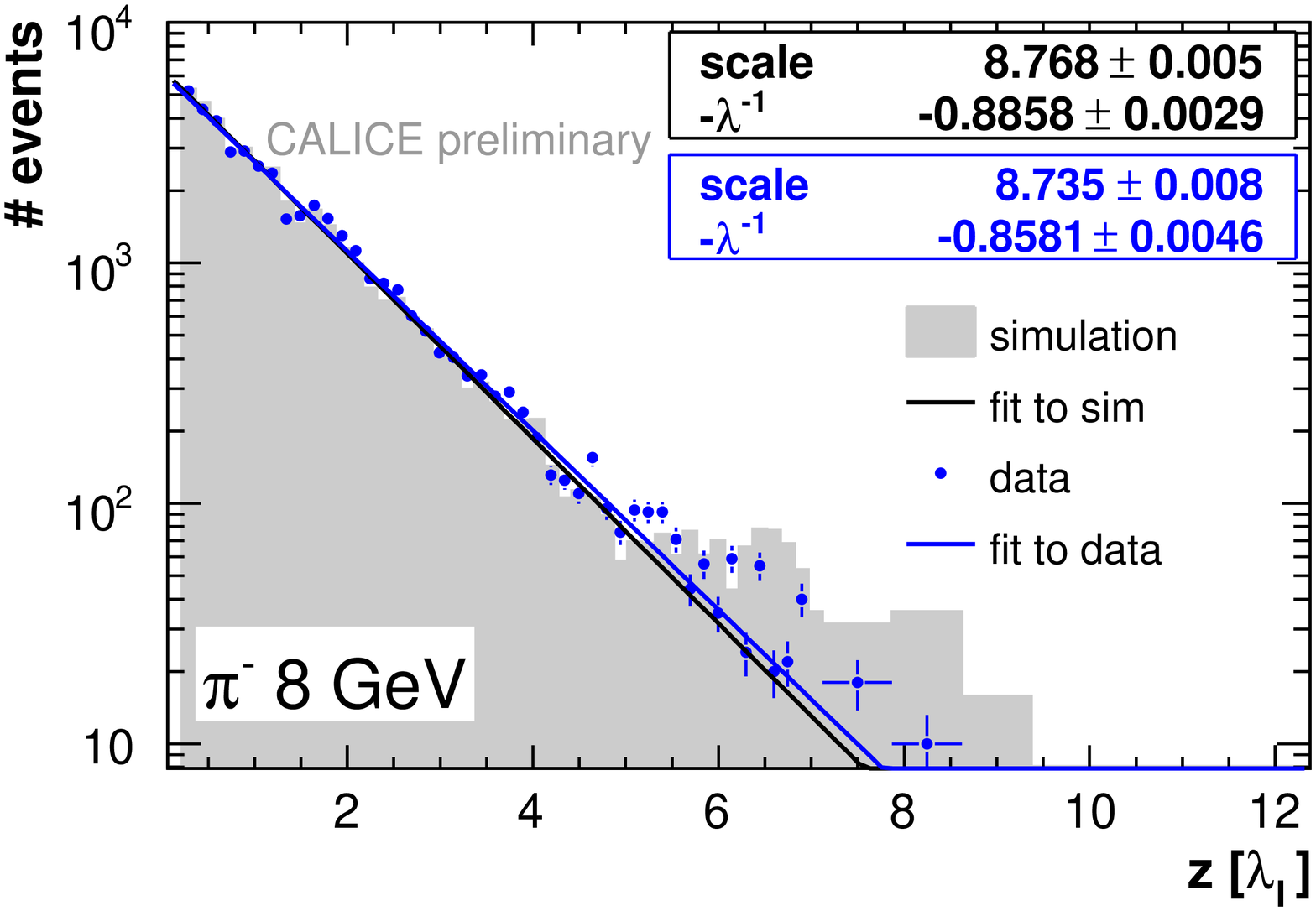}
   \vspace{-0.55cm}
   \caption{Left panel: The energy resolution to hadron-induced showers 
            is shown for both the data and MC using the 
            physics model lists QGSP\_BERT and LHEP, open squares and 
            triangles, respectively. The combined HCAL and TCMT 2006 
            CERN data were used here. 
            Right panel:
            The exponential depth distribution of the first nuclear 
            interaction in showers is measured versus the 
            nuclear interaction length for $8$ GeV $\pi^-$ mesons, 
            and compared with MC simulations using the physics model 
            list LHEP. The data shown here were collected in 2007, 
            combining the information from both the HCAL and  
            TCMT devices rotated at $30^o$ with respect to the 
            beam. The ECAL was displaced from the beam line.}
   \label{fig:HadReso}
   \vspace{-0.45cm}
\end{figure}
\section{Analysis of hadron data}
\label{sec:hadrons}
\vspace{-0.2cm}
The understanding of electromagnetic showers in the HCAL was 
shown to be reasonably enough for a preliminary analysis of
hadronic showers. The high granularity of the HCAL calorimeter 
allows the investigation of the longitudinal and lateral shower 
profiles with unprecedent precision. Different Monte Carlo physics 
model lists are available~\cite{MC}, and provide different predictions 
for the hadronic shower development, which can be constrained, 
in principle, by the precise results of the HCAL.
As a first comparison, simulations are presented
considering only the physics model lists which show the largest 
discrepancy among the investigated models, i.e., {QGSP\_BERT}
and {LHEP}~\cite{CALICE}. The results shown here were obtained using the 
above mentioned calibration procedures, and the latest MC digitisation. 

The energy resolution of the response to pion beams is shown in the left 
panel of Fig.~\ref{fig:HadReso}, and compared to MC simulations,
with the Birks' law included~\cite{MC}.  
Being the positron data analysis not fully understood yet, models 
cannot be constrained at this stage of the analysis.
%

\begin{figure}[t!]
   \vspace{-0.35cm}
   \includegraphics[width=0.50\columnwidth, height=5.6cm]
        {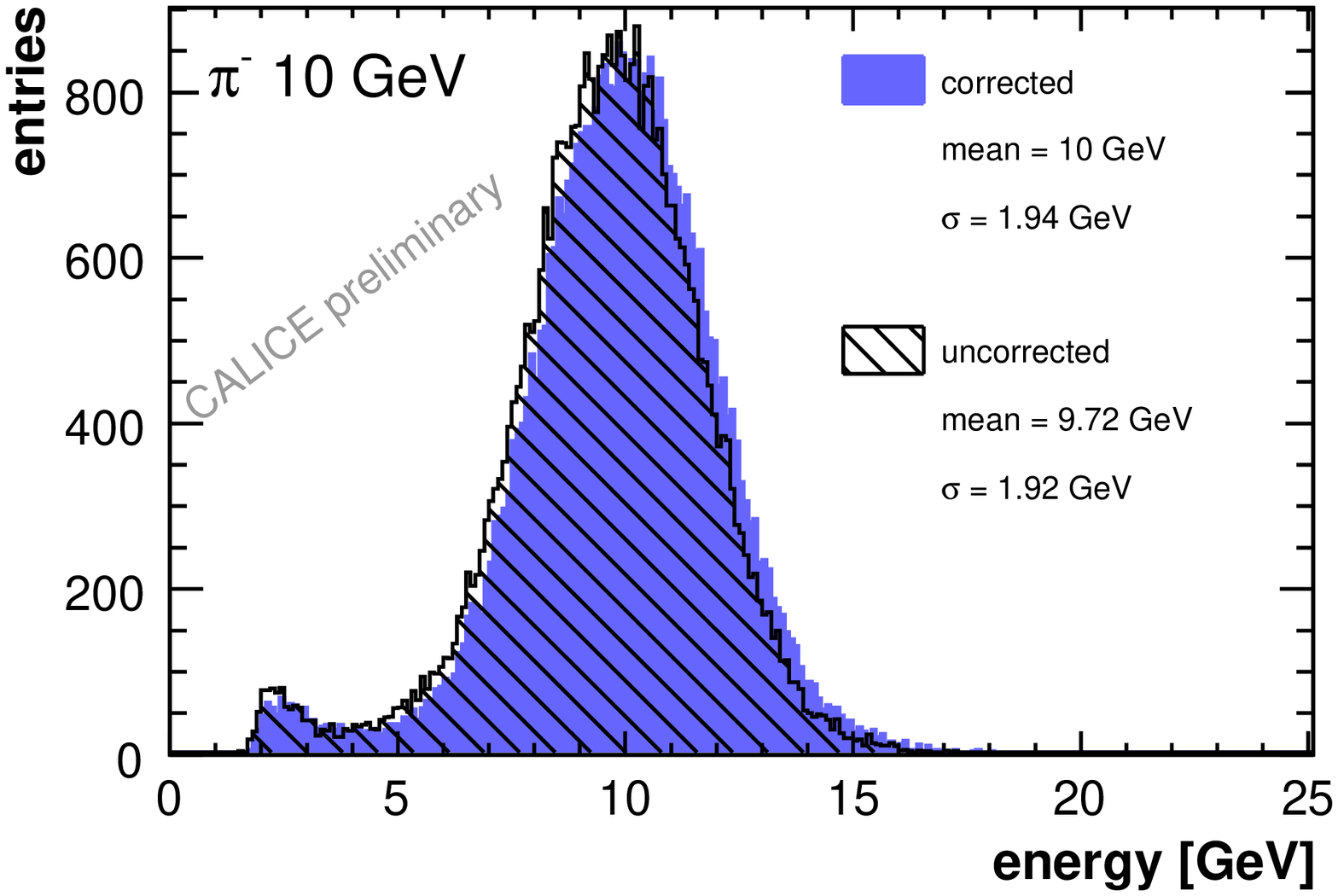}
   \hspace{-0.2cm}
   \includegraphics[width=0.55\columnwidth, height=6.cm]
        {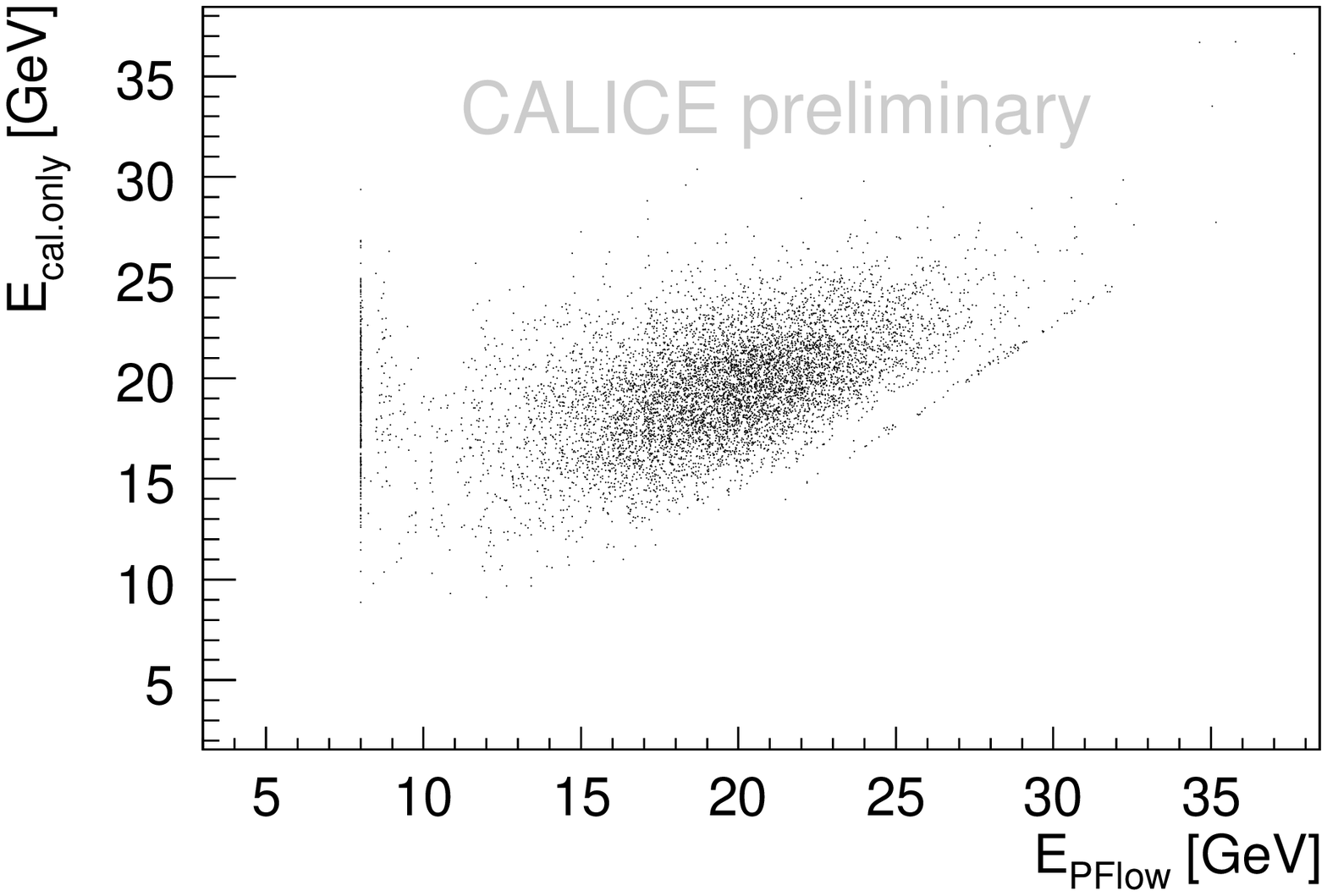}
   \vspace{-0.9cm}
   \caption{Left panel: Example of energy spectrum correction for leakage 
            effects. The same data used for the pion interaction length 
            analysis are shown here. 
            Right panel:
            Reconstructed energy sum for two pion-induced showers
            using only the calorimeter information ($E_{cal. only}$), and
            additionally applying a particle flow like approach ($E_{PFlow}$).}
   \label{fig:HadLeakage}
   \vspace{-0.3cm}
\end{figure}

The fluctuations in hadronic showers development are larger than 
what is observed in electromagnetic showers. The high 
longitudinal granularity of the HCAL allows the investigation 
of shower profiles with respect to the shower starting point.  
An exponential fit to the depth distribution of the first 
interaction in showers as a function of the distance $z$ (in nuclear 
interaction length $\lambda_I$ units) from the start of the calorimeter
was performed, and the interaction length for $\pi^-$ mesons was 
extracted~\cite{CALICE}, right panel of Fig.~\ref{fig:HadReso}. 
Comparing the reconstructed shower energy with the corresponding known 
impinging beam energy, the leakage effects in the HCAL could be 
measured with respect to the shower start location in the calorimeter, 
and corrected for~\cite{CALICE}, left panel of Fig.~\ref{fig:HadLeakage}. 

The full investigation of the PFLOW approach requires also neutral 
test-beam data, at the moment not present in the accumulated data
sets. 
Nevertheless, the proof-of-principle of the method can be investigated
merging data for hadron showers induced by pions at two different beam 
energies, thus simulating events with multiple showers. 
Showers were then reconstructed, and, for the first time using real 
data~\cite{CALICE}, the energy sum of the respective obtained two 
clusters was compared with the reconstructed energy sum obtained fixing, 
in a PFLOW like approach, one shower energy to the 
corresponding known beam energy (in a real experiment 
provided by the tracking system), right panel of Fig.~\ref{fig:HadLeakage}.
Using data with larger shower separation is expected to improve the 
energy resolution obtained using the PFLOW approach.
%
%

%
%
\begin{footnotesize}

\end{footnotesize}

\end{document}